\documentclass[a4paper,12pt]{iopart}
\usepackage[utf8]{inputenc}
\usepackage[T1]{fontenc}

\usepackage{amsmath,amssymb}
\usepackage{microtype}
\usepackage{array}
\usepackage{graphicx}
\usepackage[rgb]{xcolor}
\usepackage[numbers,sort&compress]{natbib}
\usepackage{hyperref}
\hypersetup{colorlinks,linkcolor=blue,citecolor=blue,urlcolor=blue}
\def\equationautorefname~#1\null{Eq.~(#1)\null}

\newcommand{\braket}[1]{\ensuremath{\left\langle{#1}\right\rangle}}

\begin{document}

\title[]{Comparing nonlinear optomechanical coupling in membrane-in-the-middle and single-cavity optomechanical systems}
\author{Roel Burgwal \textsuperscript{1,2}}
\author{Javier del Pino \textsuperscript{2}}
\author{Ewold Verhagen \textsuperscript{1,2}}
\address{\textsuperscript{1} Department of Applied Physics and Institute of Photonic Integration, Eindhoven University of Technology, P.O. Box 513, 5600 MB Eindhoven, The Netherlands}
\address{\textsuperscript{2} Center for Nanophotonics, AMOLF, Science Park 104, 1098 XG Amsterdam, The Netherlands}
\eads{burgwal@amolf.nl}


	\begin{abstract}
	In cavity optomechanics, nonlinear interactions between an optical field and a mechanical resonator mode enable a variety of unique effects in classical and quantum measurement and information processing. Here, we describe nonlinear optomechanical coupling in the membrane-in-the-middle (MIM) setup in a way that allows direct comparison to the intrinsic optomechanical nonlinearity in a standard, single-cavity optomechanical system. We find that the enhancement of nonlinear optomechanical coupling in the MIM system as predicted by Ludwig et al.~\cite{Ludwig2012} is limited to the degree of sideband resolution of the system. Moreover, we show that the selectivity of the MIM system of nonlinear over linear transduction has the same limit as in a single cavity system. These findings put constraints on the experiments in which it is advantageous to use a MIM system. We discuss dynamical backaction effects in this system and find that these effects per cavity photon are exactly as strong as in a single cavity system, while allowing for reduction of the required input power. We propose using the nonlinear enhancement and reduced input power in realistic MIM systems towards parametric squeezing and heralding of phonon pairs, and evaluate the limits to the magnitude of both effects. 
\end{abstract}

%
%
%
%
	%
	\maketitle
	\DeclareGraphicsExtensions{.pdf,.png,.jpg}
	
	\section{Introduction}
	Cavity optomechanics enables a wide variety of control over either optical or mechanical degrees of freedom by exploiting radiation pressure interactions. Using an effectively linear optomechanical coupling, many celebrated effects have been demonstrated, such as optical sideband cooling through dynamical backaction~\cite{Arcizet2006,Chan2011}. On the other hand, \emph{nonlinear} optomechanical interaction has been recognised as a potential resource to generate nonclassical optical and mechanical states~\cite{Rabl2011,Nunnenkamp2011}. In particular, quadratic optomechanical coupling, for which optical eigenmode frequencies scale with the square of mechanical displacement, offers several quantum applications such as a phonon quantum non-demolition (QND) measurements~\cite{Braginsky1980,Thompson2008}, squeezing of optical and mechanical modes~\cite{Nunnenkamp2010}, the observation of phonon shot noise~\cite{Clerk2010}, sub-Poissonian phonon lasing~\cite{Lorch2015}, controlled quantum-gate operations between flying optical or stationary phononic qubits~\cite{Stannigel2012} and nonclassical state generation through measurement~\cite{Brawley2016}. Additionally, there are also classical applications, such as a 2-phonon analogue of optomechanically-induced-transparency~\cite{Huang2011}. Moreover, systems that feature quadratic coupling offer new ways to let mechanical modes interact with quantum two-level systems~\cite{Cotrufo2017,Ma2020}
	
	Even the simplest optomechanical systems, where a single cavity is parametrically coupled to a mechanical resonator, feature nonlinear interaction between the optical and mechanical degrees of freedom described by the Hamiltonian:
	\begin{equation} \label{eq:single_hamiltonian}
	\hat{H} = \Omega_m \hat{b}^\dagger \hat{b} + \left[\omega_c  - g_0 (\hat{b}^\dagger + \hat{b})\right]\hat{a}^\dagger \hat{a},
	\end{equation}
	where $\Omega_m$ and $\omega_c$ are the mechanical and optical mode frequencies, respectively, $g_0$ is the single-photon optomechanical coupling rate, $\hat{a}$ and $\hat{b}$ are the optical and mechanical annihilation operator, respectively, and we set $\hbar=1$~\cite{Aspelmeyer2014}. For nonlinear effects to be appreciable for quantum-level motion, however, one requires the so-called single-photon strong coupling (SPSC) regime $g_0/\kappa > 1$, where $\kappa$ is the optical mode decay rate~\cite{Rabl2011,Nunnenkamp2011}. As this SPSC condition is inaccessible in solid-state optomechanical systems, most experiments use large coherent optical fields, that effectively linearise the linear interaction. It was recognised that special forms of nonlinear optomechanics could be achieved in multimode systems~\cite{Thompson2008,Jayich2008,Ludwig2012}. The so-called membrane-in-the-middle (MIM) system consists of two cavities coupled through optical tunnelling at rate $J$. If a mechanical mode, e.g. that of a highly reflective membrane that separates the two cavities, alters the cavity lengths with equal magnitude but opposite sign, the frequencies of the optical supermodes depend on the square of displacement to lowest order. Such quadratic coupling is described by terms $\propto (b^\dagger+b)^2 a^\dagger a$ in the Hamiltonian, whose magnitude scales inversely with $J$~\cite{Thompson2008,Jayich2008}. Here $a$ refers to one of the optical supermodes. 
	
	MIM systems were realised in Fabry-Perot cavities~\cite{Thompson2008,Karuza2013}, nanoscale platforms that include ringresonators~\cite{Hill2013} and photonic crystals~\cite{Paraiso2015}, ultracold atom systems~\cite{Purdy2010} and levitated nanosphere platforms~\cite{Bullier2020}. The development of large quadratic optomechanical coupling has also inspired closely related designs ~\cite{Doolin2014,Kaviani2015,Hauer2018}.
	
	\begin{figure}
		\centering
		\includegraphics[width=\linewidth]{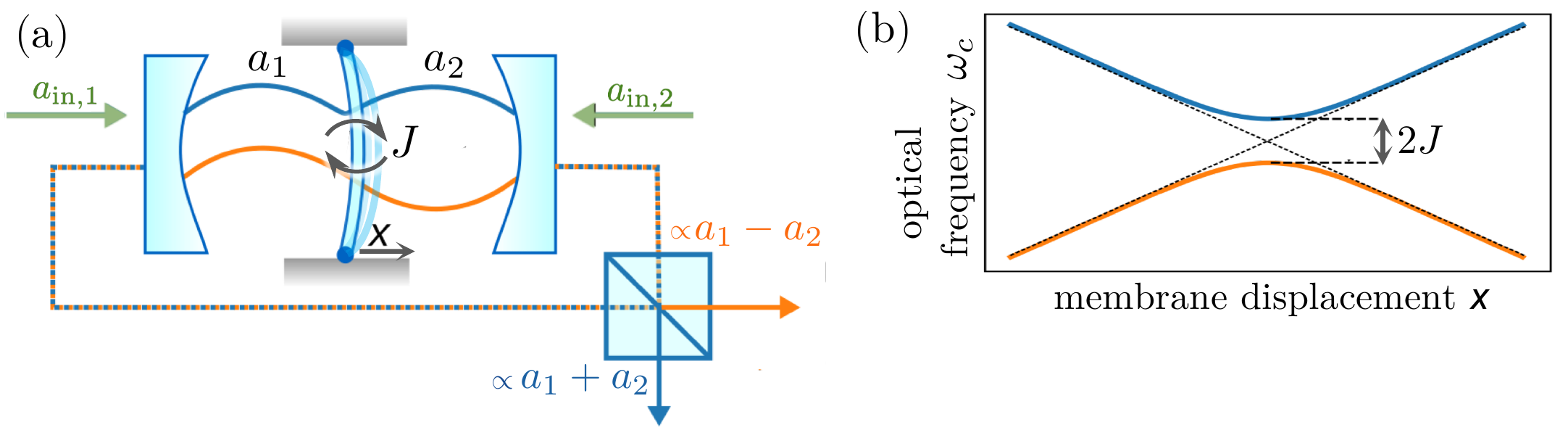}
		\caption{(a) The optomechanical membrane-in-the-middle (MIM) system, consisting of two coupled optical mode which both couple to one mechanical resonator. Note the use of both ports for input and output. (b) optical eigenfrequencies for varying oscillator position $x$.}
		\label{fig:fig1}
	\end{figure}
	
	Although optomechanical interaction in the MIM system is often described by only the quadratic interaction ~\cite{Nunnenkamp2010,Huang2011,Jiang2016,Xie2016,Liao2013,Xu2020}, it is generally an insufficient description. In addition to quadratic coupling, the mechanical mode also creates linear cross-coupling between the two optical supermodes~\cite{Biancofiore2011,Cheung2011}, allowing quantum vacuum fluctuations to excite the mechanical resonator and precluding phonon QND measurements, that become limited to the SPSC condition~\cite{Miao2009}. Moreover, when the frequency splitting of the optical supermodes is comparable to the mechanical frequency, i.e. $2J-\Omega_m\ll2J$, quadratic optomechanical coupling is resonantly enhanced ~\cite{Heinrich2011,Ludwig2012,Stannigel2012,Liao2014a,Liao2015,Lorch2015}, an effect which is also not captured in a model in which quadratic coupling is explained through the interaction of a mechanical mode with a single optical mode at an avoided crossing of optical supermodes (\autoref{fig:fig1}b). This picture is only applicable in the regime where mechanical motion can be regarded as quasi-static, i.e. $\Omega_m \ll 2J$. A general description of MIM system dynamics that extends beyond these constraints is still missing. Moreover, it is an open question how strong quadratic coupling in the MIM system can be made to be, and how that compares to the nonlinear interaction in a single cavity of similar size and optomechanical properties. Having such a description is useful in determining how quadratic optomechanical coupling can be achieved in general systems, for either quantum or classical applications, and to identify applications in the regime of weak optomechanical coupling $g_0<\kappa$ that is experimentally widely relevant.
	
	In this work, we aim to provide an intuitive description of optomechanical dynamics of the MIM system that is valid for arbitrarily small optical mode spacings and use it to describe its unique features and limitations. We quantify the strength of linear and nonlinear processes through the amplitude of the intracavity sidebands at $\pm\Omega_m$ and $\pm 2\Omega_m$, respectively, which give the strength of transduction of the mechanical mode onto the optical field, but also determine the dynamical backaction effect~\cite{Jayich2008}. These amplitudes also provide useful information about the system in the quantum regime: as Stokes and anti-Stokes (inelastic) scattering are associated with phonon generation, the $\omega_L-2\Omega_m$ sideband amplitude controls the rate of generation of pairs of phonons. Second-order sideband amplitudes determine the imprecision on a measurement of $\hat{x}^2$~\cite{Leijssen2017}, and linear sideband amplitudes determine the linear backaction of such a measurement. In constraining the discussion to only these two pairs of sidebands, we assume small cavity frequency fluctuations due to motion, $\sqrt{\langle \hat{x}^2 \rangle}g_0 < \kappa$. Because current optomechanical devices are not in the SPSC regime, this holds for most devices, although exceptions with large $g_0$ and thermal excitation break this condition ~\cite{Leijssen2017}. Indeed, various applications of quadratic coupling rely on this more practically reached coupling regime~\cite{Clerk2010,Lorch2015,Cotrufo2017,Ma2020}. Next, we revisit the dynamical backaction that the mechanical resonator experiences. Our analysis underlines that the apparent quadratic coupling in the MIM system is due to the intrinsic optomechanical nonlinearity. In particular, we see that linear transduction (i.e. the $\pm\Omega_m$ sidebands) can not be entirely suppressed and is related in size to quadratic ($\pm 2\Omega_m$ sidebands) transduction in the same way as in a single cavity system. Importantly, we show that the magnitude of the nonlinear enhancement, with respect to a single (uncoupled) optomechanical cavity, for the optimal condition of $2J=\Omega_m$ is limited to the sideband resolution $2\Omega_m/\kappa$. By describing dynamical backaction with the same approach, we put previous results on the optical spring shift and heating in a MIM system~\cite{Paraiso2015,Lee2015} in a new perspective, the most critical point being that the backaction per intracavity photon is equal in size in the two different systems. However, the multimode nature of the MIM system can be exploited to reduce the input power significantly~\cite{Dobrindt2010}. We discuss a two-tone parametric driving scheme in a MIM system that also has a reduced threshold power compared to a single cavity. Finally, we propose a scheme that exploits the enhanced nonlinearity in the MIM system to herald nonclassical 2-phonon states and works with a lower cavity occupation.
	
	This paper is organised as follows. In \autoref{sec:model}, we introduce the model and analytical results for the linear and quadratic optical transduction sidebands. We analyse these results in \autoref{sec:transd} and trace out links with existing approaches to the quasi-static ($2J\gg\Omega_m$) regime.  We subsequently focus on the enhancement of nonlinear effects that is expected in the resonant ($2J\approx\Omega_m$) case and describe the upper bounds for this nonlinearity. Next, in \autoref{sec:apps}, we estimate dynamical backaction by calculating the optically-induced changes in the mechanical response in the MIM system. We discuss these results and how, in the case of a two-tone parametric driving scheme, the MIM system can be exploited to reduce required driving power. Finally, we discuss also how the description of the MIM system in this paper might shed light on quadratic coupling in general optomechanical systems. 
	
	\section{Model and method}\label{sec:model}
	
	\subsection{First and second order sidebands in a single cavity}
	We begin by revisiting the linear and intrinsic nonlinear optomechanical coupling that occurs in single cavity optomechanical system. The optical mode couples to an external input of output field with rate $\kappa_\mathrm{ex}$. The mechanical dissipation rate is $\Gamma_m$. Starting from the Hamiltonian of \autoref{eq:single_hamiltonian}, moving to a frame rotating at the laser drive frequency $\omega_L$ and introducing the laser detuning $\Delta=\omega_L-\omega_c$, the quantum Langevin equations can be derived. These govern the dynamics of the operators in the open quantum system~\cite{Aspelmeyer2014} and read 
	\begin{subequations}
		\begin{align}
		\dot{\hat{a}}=& -\frac{\kappa}{2}\hat{a}+i(\Delta+g_0\hat{x})\hat{a} + \sqrt{\kappa_\mathrm{ex}}\hat{a}_\mathrm{in}+\sqrt{\kappa_0}\hat{f}_\mathrm{in}\\
		\dot{\hat{x}}=&\Omega_m\hat{p}\\
		\dot{\hat{p}}=&-\Omega_m\hat{x}-\Gamma_m\hat{p}+g_0\hat{a}^{\dagger}\hat{a} - \frac{\hat{F}_\mathrm{in}}{m\Omega_m x_\mathrm{zpf}}+\sqrt{\Gamma_m}\hat{P}_\mathrm{in},
		\end{align}
	\end{subequations}
	where we have used the unitless mechanical position and momentum operators, $\hat{x}=\frac{1}{\sqrt{2}}(\hat{b}^\dagger + \hat{b})$ and $\hat{p}=\frac{i}{\sqrt{2}}(\hat{b}^\dagger - \hat{b})$, respectively. We have introduced input fields $\hat{a}_\mathrm{in},\hat{f}_\mathrm{in}$, for the optical input field through the external channel and quantum fluctuations that enter the system through instrinsic decay, with rates $\kappa_\mathrm{ex}$ and $\kappa_0$, respectively, fulfilling $\kappa_0+\kappa_\mathrm{ex}=\kappa$, where $\kappa$ is the total decay rate. The field $\hat{P}_\mathrm{in}$ introduces mechanical fluctuations associated with coupling to a thermal bath whereas $\hat{F}_\mathrm{in}$ accounts for coherent mechanical drive fields ($\hat{H}_\mathrm{d}=-\hat{x}\hat{F}_{\mathrm{in}}/(m \Omega_m x_\mathrm{zpf})$). Also, $x_\mathrm{zpf} \equiv \sqrt{1/(2m\Omega_m)}$ is the mechanical zero point motion for the mechanical oscillator with effective mass $m$. In our calculations, we reduce these equations to the semiclassical -nonlinear- equations of motion in the mean-field approximation $\braket{\hat{x}\hat{a}}\simeq x a$,  denoting $\braket{\hat{a}}=a$ and $\braket{\hat{x}}=x$. Assuming no external mechanical forces ($\braket{\hat{F}_\mathrm{in}}=0$) and incoherent (e.g. thermal) input fluctuations,  $\braket{P_\mathrm{in}}=0,\braket{f_\mathrm{in}}=0$, and we arrive to:
	\begin{subequations}
		\begin{align}\label{single_opticalEOMS}
		\ddot{x} &= -\Omega_m^2 x - \Gamma_m\dot{x} + \Omega_m g_0  |a|^2, \\
		\dot{a} &= i(\tilde{\Delta} + g_0 x)a+ \sqrt{\kappa_\mathrm{ex}} a_{\mathrm{in}}.
		\end{align}
	\end{subequations}
	Here we, for convenience, absorbed the optical decay rate as an imaginary part of the complex detuning $\tilde{\Delta}$: $\kappa = 2 \mathrm{Im}(\tilde{\Delta})$ First, we find steady-state solutions:
	\begin{subequations}
		\begin{align}
		\bar{a} &= i\frac{\sqrt{\kappa_\mathrm{ex}}}{\bar{\Delta}} a_\mathrm{in}, \\
		\bar{x} &= \frac{g_0}{\Omega_m}|\bar{a}|^2.
		\end{align}
	\end{subequations}
	Here, $\bar{\Delta} = \tilde{\Delta} + g_0 \bar{x}$, which still contains $\bar{x}$. However, we will assume that the optical power is limited such that the static displacement of the resonator is much smaller than the linewidth, $g_0 \bar{x} \ll \kappa$, such that $\bar{\Delta} \approx \tilde{\Delta}$. This sets an upper limit for a few hundred intracavity photons in photonic crystal systems~\cite{Safavi-Naeini2011a},  while for other system it is much less restricting.
	
	We will evaluate the optical sidebands created by coherent mechanical motion of a specific amplitude $X_0$, described by $x = \bar{x} + X_0 \cos(\Omega_m t)$. For the optical field, we look for a perturbative solution of the form~\cite{Heinrich2011}:
	\begin{equation} 
	a(t)=\bar{a}+\sum_{\zeta=\pm}A_{\zeta}^{(1)}e^{i\zeta\Omega_{m}t}+A_{\zeta}^{(2)}e^{i\zeta2\Omega_{m}t}.
	\end{equation}
	By collecting terms in the mean-field EOM with the same time dependence, we can solve for the first-order coefficients:
	\begin{equation} \label{eq:singlecav_firstsb}
	A_{\pm}^{(1)} = \frac{g_0\bar{a}}{\pm \Omega_m - \bar{\Delta}} \frac{X_0}{2}.
	\end{equation}
	And, using this result, we can also retrieve second-order coefficients
	\begin{equation} \label{eq:singlecav_secondsb}
	A_{\pm}^{(2)} = \frac{g_0 A_{\pm}^{(1)}}{\pm 2\Omega_m - \bar{\Delta}} \frac{X_0}{2} = \frac{g_0^2 \bar{a}}{(\pm 2\Omega_m-\bar{\Delta})(\pm \Omega_m-\bar{\Delta})} \big(\frac{X_0}{2}\big)^2.
	\end{equation}
	
	In the approach we take above, the hierarchy of higher-order sidebands has been truncated assuming the cavity resonance frequency shift because of mechanical motion is negligible compared to the optical linewidth, i.e. $g_0|x|<\kappa$, in which case every higher-order sideband can be treated as a perturbation of the previous.
	
	\subsection{Interaction and sidebands in the MIM system}
	Having applied our approach to single cavities, we now move to the MIM system. Our starting point is the standard Hamiltonian of the MIM system in the rotating frame of an input laser field detuned from two optical modes by $\mathrm{Re}\tilde{\Delta}_i=\omega_L-\omega_{c,i}$, with loss rates $2\mathrm{Im}\tilde{\Delta}_i=\kappa_i$ that are coupled to a single mechanical membrane, displaced from the equilibrium position by $\hat{x}$. In the basis of the physical cavities with annihilation operators $\hat{a}_i$ ($i=\{1,2\}$) the system is governed by the Hamiltonian 
	\begin{equation}
	\hat{H}=\Omega_m \hat{b}^\dagger \hat{b}+\hat{H}_\mathrm{OM}+\hat{H}_{J} + \sum_i\hat{H}_{\kappa_i},\label{eq:system_H}
	\end{equation}
	where optomechanical coupling reads
	\begin{equation} \label{baremodeeqs}
	\hat{H}_\mathrm{OM} = (\Delta_1 - g_{0,1}\hat{x})\hat{a}_1^\dagger \hat{a}_1 + (\Delta_2  +g_{0,2}\hat{x})\hat{a}_2^\dagger \hat{a}_2,
	\end{equation}
	and the optical inter-cavity coupling is characterized by
	\begin{equation}
	\hat{H}_{J} =  - J(\hat{a}_1^\dagger\hat{a}_2 + \hat{a}_2^\dagger \hat{a}_1),
	\end{equation}
	where $J$ is the rate of inter-cavity coupling. Coupling to input/output channels via Hamiltonians $\hat{H}_{\kappa_i}$ is assummed to occur to separate environments, (e.g. single-mode waveguides)  with rates $\kappa_\mathrm{ex,i}$. Because the optical cavities are coupled, \autoref{eq:system_H} can be expressed in terms of the optical supermodes that arise. In conditions of equal cavity frequency $\Delta_1=\Delta_2 \equiv \Delta$ and optomechanical coupling $g_{0,1}=g_{0,2}\equiv g_0$, these are given by $\hat{a}_{e,o}=(\hat{a}_1\pm\hat{a}_2)/\sqrt{2}$. These supermodes are also depicted in \autoref{fig:fig1}a.  
	In this basis $\hat{H} = \sum_{\eta=e,o}\omega_\eta\hat{a}_\eta^\dagger \hat{a}_\eta + \hat{H}_\mathrm{OM} + \Omega_m\hat{b}^{\dagger}\hat{b}$ with $\omega_{e,o}=\Delta\mp J$, with an optomechanical interaction:
	\begin{equation}
	\hat{H}_\mathrm{OM}  =  -g_0 \hat{x}(\hat{a}_e^\dagger \hat{a}_o + \hat{a}_o^\dagger \hat{a}_e).\label{supermodeeqsb}
	\end{equation}
	Here, we want to emphasize the fact that optomechanical coupling has now become cross-mode, i.e. the Hamiltonian contains terms $\propto \hat{x}\hat{a}_e^\dagger \hat{a}_o$, whereas it previously contained self-mode terms, e.g. $\propto \hat{x}\hat{a}_1^\dagger \hat{a}_1$.
	
	The frequencies of these optical supermodes can be found by treating this mechanical position as a quasi-static parameter analogous to the Born-Oppenheimer approximation of molecular physics ($\hat{x}\mapsto x$). This is only valid for mechanical motion that is slow with respect to the optical coupling rate, or $J \gg \Omega_m$, which is not true for a number of experimental implementations ~\cite{Thompson2008,Grudinin2010,Hill2013}. Using this approximation allows for diagonalization of the system Hamiltonian in \autoref{eq:system_H}~\cite{Jayich2008}, yielding the $x$-dependent eigenfrequencies in  \autoref{fig:fig1}b. Still assuming equal frequency of both optical cavities, this  dependence is approximately quadratic and given by $ \omega^{\mathrm{ad.}}_{e,o}(x) \simeq \Delta \mp (J + g_0^{(2)}x^2)$, or, equivalently, the effective quadratic coupling Hamiltonian 
	\begin{equation}\label{eq:effective_quad}
	\hat{H}^{\mathrm{ad.}}=\Delta\left(\hat{a}_{e}^{\dagger}\hat{a}_{e}+\hat{a}_{o}^{\dagger}\hat{a}_{o}\right)-(J+g_0^{(2)}\hat{x}^{2})(\hat{a}_{e}^{\dagger}\hat{a}_{e}-\hat{a}_{o}^{\dagger}\hat{a}_{o}),
	\end{equation}
	with effective quadratic coupling $g_0^{(2)}= g_0^2/2J$. It is this form of the Hamiltonian that drew attention to the MIM system as a platform for strong quadratic optomechanical coupling. This adiabatic limit, however, breaks down as optical Rabi oscillations occur at scales that compare with mechanical oscillations, i.e. where the supermode splitting approaches the mechanical frequency ($2J\approx \Omega_m$). In this limit, optical and mechanical degrees of freedom need to be treated on the same footing, via numerical methods or effective Hamiltonians that are perturbative in $g_0$~\cite{Yanay2017,Ludwig2013}. Moreover, as described in the introduction, it was quickly recognised that this effective Hamiltonian does not fully describe the system, because the linear cross-mode coupling is no longer included~\cite{Miao2009,Yanay2016}. 
	
	In order to provide a more complete, while still intuitive picture of the MIM system dynamics that naturally covers adiabatic and resonant regimes, we apply the same perturbative approach as with the single cavity to the full model in \autoref{baremodeeqs}. Our mean-field equations of motion are:
	\begin{subequations}
		\begin{align}\label{eq:opticalEOMS}
		\Ddot{x} &= -\Omega_m^2 x - \Gamma_m\dot{x} + \Omega_m(g_{0,1}|a_1|^2 - g_{0,2}|a_2|^2) +\frac{F_\mathrm{in}}{mx_\mathrm{xpf}},\\
		\dot{a}_1 &= i(\tilde{\Delta}_1 + g_{0,1} x)a_1 + iJa_2 + \sqrt{\kappa_\mathrm{ex,1}} a_{\mathrm{in},1}, \\
		\dot{a}_2 &= i(\tilde{\Delta}_2 - g_{0,2} x)a_2 + iJa_1 + \sqrt{ \kappa_\mathrm{ex,2}} a_{\mathrm{in},2}.
		\end{align}
	\end{subequations}
	Here $m$ stands for the effective oscillator mass and the optical decay rates $\kappa_i=-2\mathrm{Im}{\tilde{\Delta}_i}$ are included in the complex detunings $\tilde{\Delta}_i$. We have added the term $\propto F_\mathrm{in} =\braket{\hat{F}_\mathrm{in}}$ to represent external classical forces acting on the resonator, which will be of use later on. We first find steady state values for $a_1, a_2$ and $x$:
	\begin{subequations} \label{eq:opticalSteady}
		\begin{align}
		\bar{a}_{1,2}&=i\frac{(\bar{\Delta}_{1,2}\xi_{1,2}+J\xi_{2,1})}{J^{2}-\bar{\Delta}_{1}\bar{\Delta}_{2}},\\
		\bar{x} &= \frac{g_{01}|a_1|^2 - g_{02}|a_2|^2)}{\Omega_m}.
		\end{align}
	\end{subequations}
	Where $\bar{\Delta}_i = \tilde{\Delta}_i \pm g_{0i}\bar{x}$ is the detuning to the cavity resonance that has been displaced by mean mechanical position $\bar{x}$ and incoming photon population $\xi_i = \sqrt{\kappa_\mathrm{ex} }a_{\mathrm{in},i}$. Similarly to the discussion of the single-cavity intrinsic nonlinearity, we propose an ansatz
	\begin{equation}
	a_{i}=\bar{a}_{i}+\sum_{\zeta=\pm}A_{i,\zeta}^{(1)}e^{i\zeta\Omega_{m}t}+A_{i,\zeta}^{(2)}e^{i\zeta2\Omega_{m}t}.
	\end{equation}
	We then derive explicit expressions for the first-order coefficients, 
	\begin{subequations}\label{firstsb}
		\begin{align}
		A_{1,\pm}^{(1)}    &= -\frac{X_{0}}{2}\frac{-J g_{0,2}\bar{a}_2 + (\bar{\Delta}_2\mp\Omega_m)g_{0,1} \bar{a}_1}{(\bar{\Delta}_1\mp \Omega_m)(\bar{\Delta}_2 \mp \Omega_m )-J^2},  \\
		A_{2,\pm}^{(1)}    &= \frac{X_{0}}{2}\frac{-J g_{0,1}\bar{a}_1 + (\bar{\Delta}_1\mp\Omega_m)g_{0,2} \bar{a}_2}{( \bar{\Delta}_1\mp\Omega_m )(\bar{\Delta}_2\mp\Omega_m)-J^2} ,
		\end{align}
	\end{subequations}
	as well as the second order coefficients
	\begin{subequations}\label{fullsb2}
		\begin{align} 
		A_{1,+}^{(2)}    =&-\frac{X_{0}}{2}\frac{g_{0,1}A^{(1)}_{1,+}(\bar{\Delta}_{2}-2\Omega_{m})-g_{0,2}A^{(1)}_{2,+}J}{(\bar{\Delta}_{1}-2\Omega_{m})(\bar{\Delta}_{2}-2\Omega_{m})-J^{2}},\\
		A_{1,-}^{(2)}    =&-\frac{X_{0}}{2}\frac{g_{0,1}A^{(1)}_{1,-}(\bar{\Delta}_{2}+2\Omega_{m})-g_{0,2}A^{(1)}_{2,-}J}{(\bar{\Delta}_{1}+2\Omega_{m})(\bar{\Delta}_{2}+2\Omega_{m})-J^{2}},\\
		A_{2,+}^{(2)}    =&\frac{X_{0}}{2}\frac{-g_{0,1}A^{(1)}_{1,+}J+g_{0,2}A^{(1)}_{2,+}(\bar{\Delta}_{1}-2\Omega_{m})}{(\bar{\Delta}_{1}-2\Omega_{m})(\bar{\Delta}_{2}-2\Omega_{m})-J^{2}},\\
		A_{2,-}^{(2)}    =&\frac{X_{0}}{2}\frac{-g_{0,1}A^{(1)}_{1,-}J+g_{0,2}A^{(1)}_{2,-}(\bar{\Delta}_{1}+2\Omega_{m})}{(\bar{\Delta}_{1}+2\Omega_{m})(\bar{\Delta}_{2}+2\Omega_{m})-J^{2}}.
		\end{align}
	\end{subequations}

	\section{Optomechanical transduction} \label{sec:transd}
	
	Having obtained the expressions for the sideband amplitudes for a given mechanical amplitude, we now discuss these results in the context of mechanical transduction. We begin by retrieving the results of the quasi-static model from our approach.
	
	\subsection{Recovering the quasi-static limit} \label{sec:quasistatic}
	
	\begin{figure}[t]
		\centering
		\includegraphics[width=\linewidth]{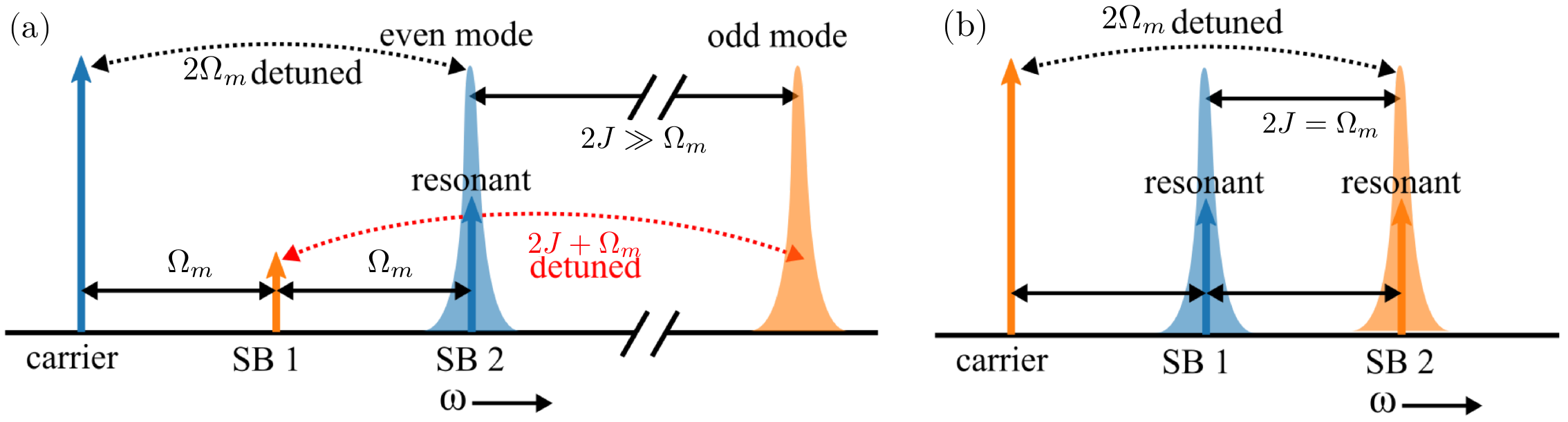}
		\caption{Schematic depiction of the frequencies of input field, sidebands and optical modes. Blue and orange colour coding indicate the even and odd optical modes, respectively. (a) In the adiabatic limit ($2J\gg \Omega_m$) and for driving of the even mode, the first sideband (SB1) is far off resonance with the odd mode, while the second sideband (SB2) is on resonance with the even mode. (b) Conversely, for $2J\simeq\Omega_m$, a doubly resonant condition can be satisfied. Results are shown for $\Delta=J+2\Omega_m$.}
		\label{fig:fig2}
	\end{figure}
	
	Here, we impose the quasi-static limit ($2J \gg \Omega_m$) in the general solutions above and assume mode splitting to be larger than the individual modes ($2J \gg \kappa_i$). Without loss of generality, we drive the input of cavity 1 close to the even optical supermode, resulting in $\bar{a}_1 \approx \bar{a}_2$ according to \autoref{eq:opticalSteady}, but such that the $2\Omega_m$ sideband is on resonance: $\bar{\Delta} = 2\Omega_m+J$ (see \autoref{fullsb2}). We will assume a sideband resolved system with $\Omega_m>\kappa$, which is the more interesting regime for the MIM, as we will discuss later.
	
	The quasi-static diagonalization approach shows that photonic eigenmodes acquire a dependence on $x$. For $\kappa_1\neq\kappa_2$, this in addition yields an effective $x$-dependent supermode decay rate (also known as \textit{dissipative} coupling~\cite{Wu2014,Yanay2016}), leading to information about $\hat{x}$ leaking from the cavity. In a similar but distinct effect, the two optical supermodes also become coupled through their dissipation into the same optical channel for $\kappa_1\neq\kappa_2$~\cite{Dobrindt2010,Yanay2016}. However, for clarity of our discussion, we will neglect both of these effects by assuming identical optical cavities ($g_{0,1}=g_{0,2}\equiv g_0$, $\Delta_1=\Delta_2\equiv\Delta$, and $\kappa_1=\kappa_2\equiv\kappa$). 
	
	Since the drive is close to supermode resonance we have $\bar{a}_2 \approx \bar{a}_1 = \bar{a}$ and the relevant first-order sideband amplitudes reduce to
	\begin{equation} \label{firstsb_simp}
	A_{1,+}^{(1)} = \frac{g_0}{\Omega_m+J-\bar{\Delta}} \frac{\bar{a}X_0}{2}=-A^{(1)}_{2,+}.
	\end{equation}
	Here we see that this first sideband amplitude has a resonance only at the \textit{even} optical mode, or for $\mathrm{Re}(\bar{\Delta}) =\Omega_m - J $. Because this resonance frequency is far from the (odd mode) input frequency (see \autoref{fig:fig2}a), first sideband generation is suppressed. This  is a signature of the inter-mode optomechanical coupling between supermodes in \autoref{supermodeeqsb}: if the even mode is populated, the mechanical mode scatters light from the carrier into the odd mode. In figure \autoref{fig:fig2}a, we illustrate this situation. In our perturbative picture, the second sidebands at $\pm 2\Omega_m$ are seen as being scattered from the first sidebands by the mechanical mode. Because of the cross-mode coupling the second sidebands are again in the even mode. For our choice of detuning, this means the positive frequency second sideband is on resonance with the \textit{even} mode and has amplitude
	\begin{align}\label{eq:2sd_adiabatic}
	A_{1,+}^{(2)} &= A_{2,+}^{(2)} = \frac{g_0A_{1,+}^{(1)}}{2\Omega_m- J-\bar{\Delta}}\frac{X_0}{2} \approx \frac{g_0^2}{2J} \frac{a}{-i\kappa/2}\left(\frac{X_0}{2}\right)^2,
	\end{align}
	which is depicted in \autoref{fig:fig2}a. Note that a quadratic optomechanical interaction, which in practice involves the adiabatic elimination of the supermode off-resonant with the input field ($\hat{a}_o$ in this case), yields the same result for the effective quadratic coupling as in the adiabatic diagonalisation (see \autoref{eq:effective_quad}), namely $g_0^{(2)} = g_0^2/2J$. We conclude that our approach gives the correct quadratic coupling found in the quasi-static approach, but now as a manifestation of the intrinsic optomechanical nonlinearities of cavities 1 and 2, as recognised by~\cite{Stannigel2012}.
	
	\subsection{Enhanced linear and quadratic transduction}
	
	We now use our model to describe general transduction in the MIM system. In particular, we show how transduction of motion to $\Omega_m$ and $2\Omega_m$ optical sidebands changes with tunnelling rate $J$ and input laser detuning $\Delta$. In doing so, we will first assume only one optical supermode is excited by the input field, even when this field is not on resonance with that mode. This assumption makes the following discussion more clear and in fact be also achieved in experiment by exciting the MIM system through both input ports with a particular relative phase. For example, using $a_{1,\mathrm{in}} = a_{2,\mathrm{in}}$ allows excitation of only the even optical mode, regardless of optical detuning.
	
	When discussing the dynamics of the MIM system, two distinct situations can be distinguished, namely, \textit{i)} a constant input power ($P_\mathrm{in}\equiv\sum_i\omega_{c,i}|a_{i,\mathrm{in}}|^2$) or \textit{ii)} a constant cavity photon number ($\bar{n}_c\equiv\sum_i|\bar{a}_i|^2$). The latter scenario allows isolating optomechanical effects, including the strength of nonlinear transduction, from purely optical cavity input effects, i.e. the enhancement of cavity occupation for a resonant input field. Moreover, cavity occupation is often the limiting factor in the experiment, due to nonlinear effects and heating~\cite{Ren2019}. However, it could also be advantageous to minimise the input power that is required to achieve a certain cavity photon number in certain scenarios. Thus, we will discuss both situations in the following.
	
\begin{figure}[t]
		\centering
		\includegraphics[width=0.8\linewidth]{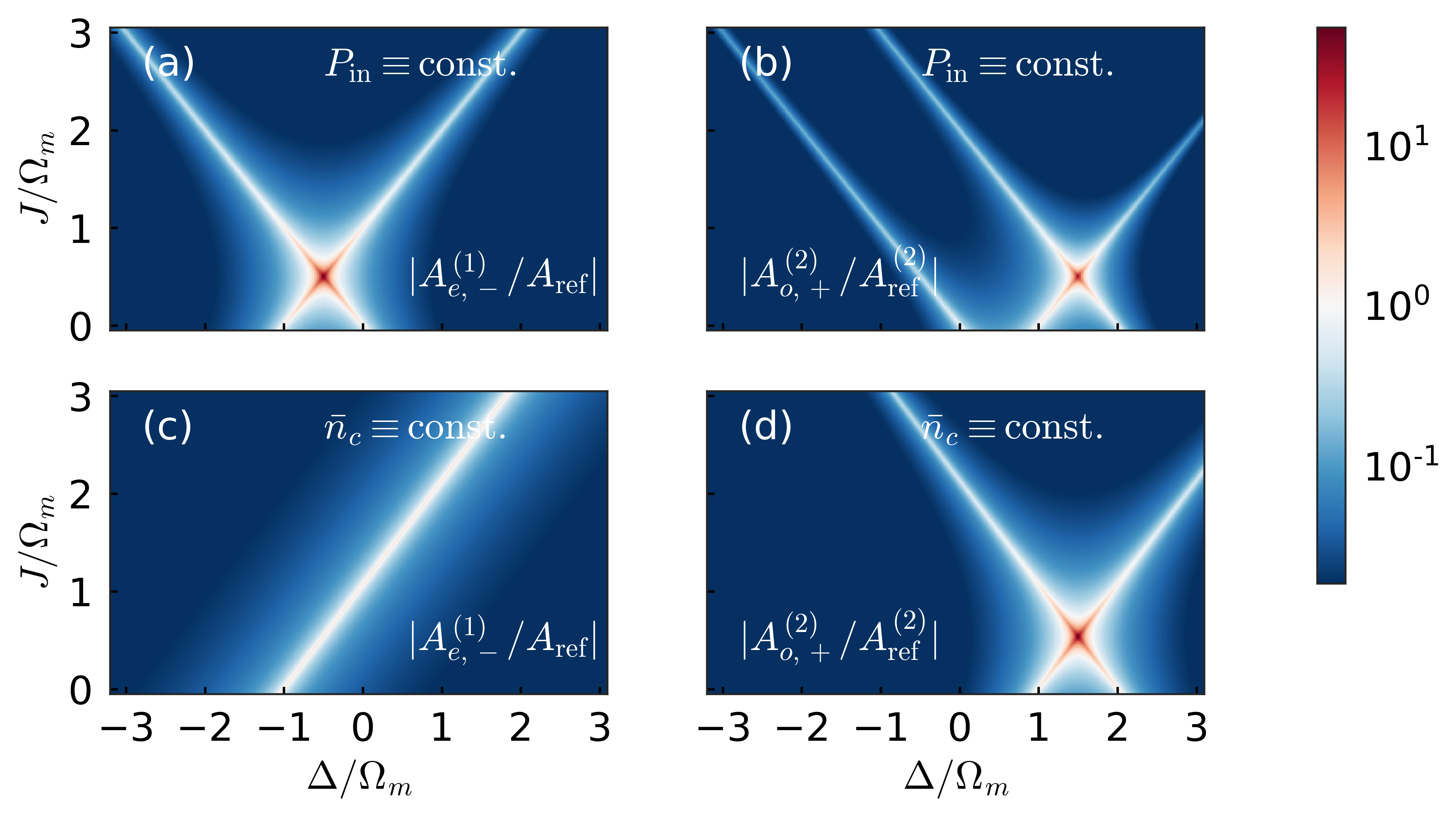}
		\caption{Mechanical transduction amplitudes as a function of laser detuning $\Delta$ and mode splitting $J$ for constant input power $P_\mathrm{in}$ (panels (a)  and (b)) or constant intracavity photon number $\bar{n}_c$ (panels (c),(d)). We depict first (left column) and second (right column) sidebands. Our colormap is chosen such that any sideband amplitude over the single cavity limits, i.e. an enhancement, is coloured red. For this plot, we used sideband resolution $\Omega_m/\kappa = 20$.}
		\label{fig:fig3}
	\end{figure}

	The amplitude of the $-\Omega_m$ and $+2\Omega_m$ sidebands of the supermodes, $A_{e,+}^{(1)}$ and $A_{o,-}^{(2)}$, for odd input ($a_{1,\mathrm{in}}=-a_{2,\mathrm{in}}$) are shown
	in \autoref{fig:fig3}  for constant $P_\mathrm{in}$ (panels \autoref{fig:fig3}a,b) and constant $\bar{n}_c$ (panels \autoref{fig:fig3}c,d). These amplitudes are defined as $A_{e,-}^{(1)} = \frac{1}{\sqrt{2}} \left( A_{1,-}^{(1)} + A_{2,-}^{(1)} \right) $ and $A_{o,+}^{(2)} = \frac{1}{\sqrt{2}} \left( A_{1,+}^{(2)} - A_{2,+}^{(2)} \right) $. The amplitudes are normalised to the optimum first sideband or second sideband amplitude that would be obtained in a single cavity for the same $P_\mathrm{in}$ or $\bar{n}_c$, which occur at $\bar{\Delta} = \pm\Omega_m$). From  \autoref{eq:singlecav_firstsb} and \autoref{eq:singlecav_secondsb}, these read
	\begin{subequations} \label{eq:refs}
		\begin{align}
		A_{+}^{(1)} (\Delta=\Omega_m)\equiv& A_\mathrm{ref} = i\frac{g_{0}\bar{a}}{\kappa}X_{0}, \\
		A_{+}^{(2)} (\Delta=\Omega_m)\equiv&A_\mathrm{ref}^{(2)} = \frac{i}{\kappa}\frac{g_{0}^{2}\bar{a}}{\Omega_{m}-i\kappa/2}\frac{X_{0}^{2}}{2},
		\end{align}
	\end{subequations}
	with $\bar{a} = \sqrt{\bar{n}_c}$ or $\bar{a} = \sqrt{ \kappa_\mathrm{ex}}a_{in}/(\kappa/2 - i\Omega_m) $ for constant $\bar{n}_c$ or $P_\mathrm{in}$, respectively. We choose to display the $+\Omega_m$ first order and $-2\Omega_m$ second order sidebands, because these show special double resonance conditions for the even mode illumination condition, as discussed below.
	
	From \autoref{fig:fig3}a, we observe strong first-order sideband generation in the even mode either when the carrier is on resonance with the odd mode ($\mathrm{Re}(\bar{\Delta})=-J$), or when the first sideband is on resonance with the odd mode ($\mathrm{Re}(\bar{\Delta})-\Omega_m=J-\Omega_m$). Where these two resonant conditions are simultaneously met, we see a resulting enhancement of first sideband generation~\cite{Dobrindt2010} and the sideband amplitude exceeds $A_\mathrm{ref}$, the largest amplitude possible in a single cavity. Moving to \autoref{fig:fig3}c, we now keep the cavity photon number $\bar{n}_c$ constant, instead of the input power. We see that the resonance of the carrier no longer results in large sideband amplitude, because we now consider a constant $\bar{n}_c$. The sideband amplitude no longer exceeds $A_\mathrm{ref}$, that of a single cavity, anywhere and we can not recognise an enhancement anymore. We conclude that the enhancement observed in \autoref{fig:fig3}a does not result from enhanced processes inside the cavity, but from a better cavity acceptance of input light.
	
	Moving to the second sideband amplitude in \autoref{fig:fig3}b, we see resonance lines that correspond to either carrier resonance or $+\Omega_m$ sideband resonance. Wherever the first positive sideband amplitude is large (not shown separately), the second sideband amplitude rises accordingly. However, an additional resonance is observed for the second-order sideband in \autoref{fig:fig3}b, where the second sideband is on resonance with the even mode ($\mathrm{Re}(\bar{\Delta})=-J+2\Omega_m$). \autoref{fig:fig3}b and \autoref{fig:fig3}d show identical dependencies, except for the line of carrier resonance ($\mathrm{Re}(\bar{\Delta})=-J$), that is not observed for constant $\bar{n}_c$.  Of special interest is the crossing of two resonance lines in the plots for quadratic transduction in \autoref{fig:fig3} (b,d), corresponding to the doubly resonant case $\mathrm{Re}(\bar{\Delta}) = 3\Omega_m/2$ and $2J=\Omega_m$. For these conditions, both the first and the second sidebands are on resonance with their respective optical mode, as we have sketched in figure \autoref{fig:fig2}b. At these points we find the strongest generation of second-order (nonlinear) sidebands, the maxima for $A_{o,+}^{(2)}$, which are larger than possible in a single cavity ($A_\mathrm{ref}^{(2)}$). Unlike with the enhanced first sideband, this effect does not disappear when considering a fixed $\bar{n}_c$. 
	
	This resonance effect has been described before by Ludwig \textit{et al.}~\cite{Ludwig2012} through a perturbative expansion of the threefold interaction between $\hat{a}_o,\hat{a}_o$ and $\hat{x}$ in \autoref{supermodeeqsb}. This leads to an effective nonlinear interaction Hamiltonian that is enhanced for $2J-\Omega_m\ll\kappa$, namely $\hat{H}_\mathrm{OM}^{\mathrm{eff}}\sim g_0^2(1/(2J-\Omega_m)+1/(2J+\Omega_m))(\hat{a}_e^\dagger\hat{a}_e-\hat{a}_o^\dagger\hat{a}_o)\hat{x}^2$. However, the magnitude of this interaction and its dependency on parameters such as $\kappa$ was not discussed. This and related works~\cite{Komar2013,Liao2015} have investigated the implications of this enhancement for specific quantum applications at the strong single-photon optomechanical coupling level ($g_0>\kappa$) and weak driving/low cavity occupation regime. In these works, it was demonstrated that the coupled cavity system had a significant advantage over a single cavity system~\cite{Stannigel2012}, but single-photon strong coupling was still needed to produce the sought-after nonclassical effects.
	
	\begin{figure*}[t]
		\centering
		\includegraphics[width=\linewidth]{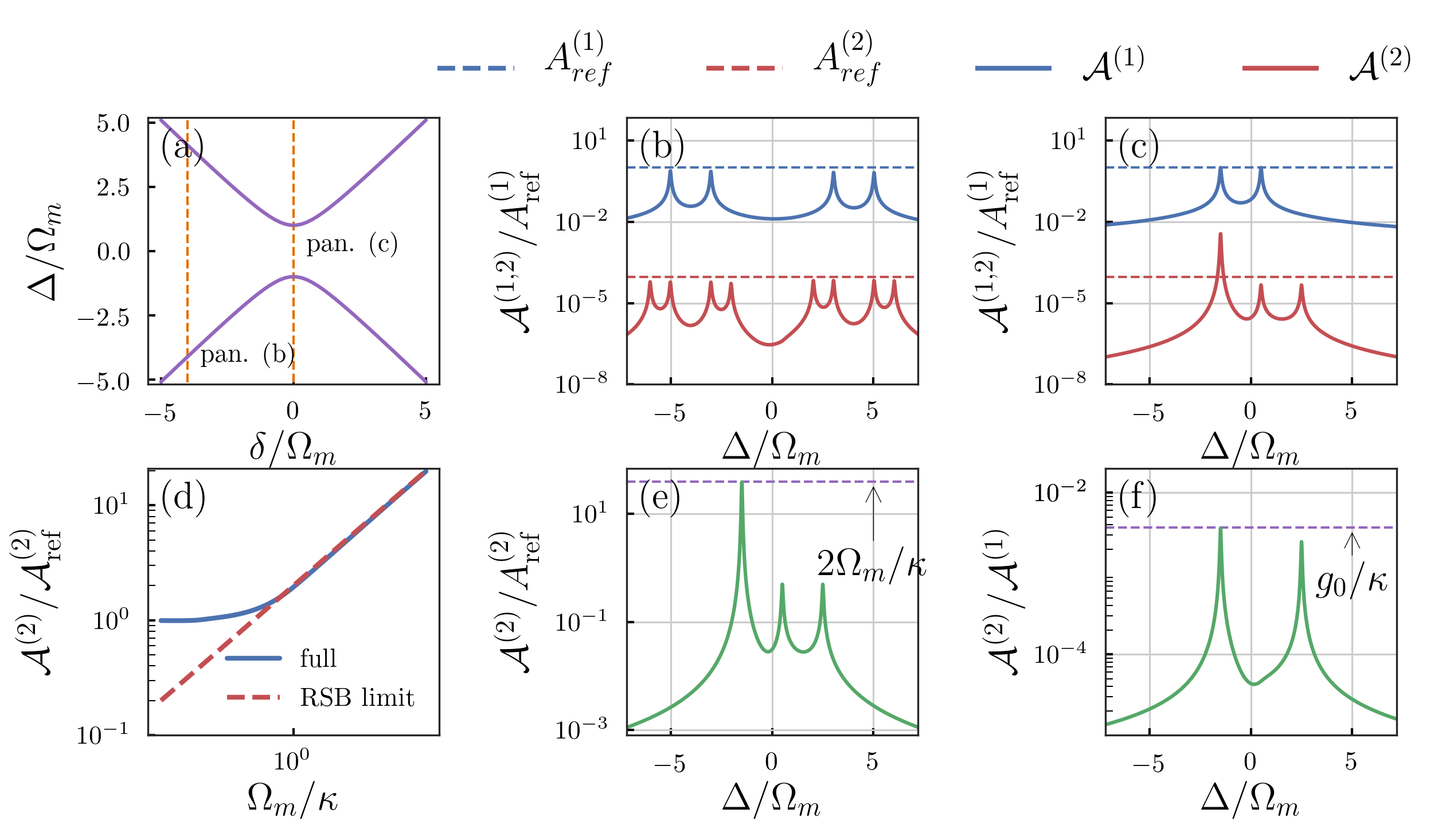}
		\caption{Different limits of mechanical transduction in a MIM system with $2J=\Omega_m$, $\Omega_m/\kappa = 20$ and ground-state motion ($X_0={x}_\mathrm{zpf}$) with optomechanical coupling $g_0/\kappa=1/250$. In (a), we show the optical eigenfrequencies as we vary the inter-cavity detuning $\delta$ and the cavities transition from uncoupled to coupled. In (b), we show linear and quadratic transduction and the corresponding single-cavity limits for the uncoupled system. We see the single-cavity limits are not exceeded. In (c), we plot the same results for the coupled system, where enhancement of nonlinear transduction is achieved. In (d), we plot the enhancement of nonlinear transduction as a function of the sideband resolution of the system, showing enhancement only exists for a sideband resolved system. In (e) we show the nonlinear enhancement for the coupled MIM system as a function of $\bar{\Delta}$. We find it peaks at $2\Omega_m/\kappa$, the degree of sideband resolution. In (c) we plot the ratio of first to second sideband amplitude as a measure of MIM system selectivity of quadratic coupling. We see it is limited by $g_0/\kappa$, just as in a single cavity system.}
		\label{fig:enhancement}
	\end{figure*}
	
	If we were to excite using even input light conditions ($a_{1,\mathrm{in}}=a_{2,\mathrm{in}}$), the roles of odd and even modes would be interchanged (not shown) and the same resonance conditions found on the $+\Omega_m$ and $-2\Omega_m$ sidebands. From a more practical perspective, using only single-port excitation of our MIM system would result in a $\Delta$-dependence convoluted with the detuning-dependent excitation of the supermodes.
	
	\subsection{Upper bounds for second sideband enhancement}
	To understand exactly what the MIM system offers over a single cavity system in terms of optomechanical nonlinearity, it is important to calculate how large the enhanced second sideband amplitude is and how this depends on system parameters. To do so, we compare optimum second sideband amplitude from \autoref{fullsb2}, i.e. at the double resonance condition described above, to optimum second sideband from a single cavity, as described in \autoref{eq:refs}. For this, we introduce a metric that combines both sidebands of the same order, namely
	\begin{subequations} \label{metric}
		\begin{align}
		\mathcal{A}^{(1)}_{s} &= |A_{s,-}^{(1)}| + |A_{s,+}^{(1)}|, \\
		\mathcal{A}^{(2)}_{s} &= |A_{s,-}^{(2)}| + |A_{s,+}^{(2)}|,
		\end{align}
	\end{subequations}
	where $s=o,e$. As shown in \ref{appendixone}, this metric is proportional to the homodyne signal amplitude at $\Omega_i$ or $2\Omega_i$ in the optimum optical quadrature. This metric can also be applied to the single-cavity case using \autoref{eq:singlecav_firstsb} and \autoref{eq:singlecav_secondsb}, to obtain the reference values $\mathcal{A}^{(1,2)}_\mathrm{ref}$.
	
	In figure \autoref{fig:enhancement}d, we plot the ratio of $\max_\Delta (\mathcal{A}^{(2)}_{e}(\Delta))$ and $\max_\Delta (\mathcal{A}^{(2)}_\mathrm{ref}(\Delta))$ for even input drive as a measure of the enhancement of nonlinearity for different values of sideband resolution $\Omega_m/\kappa$. For sideband-unresolved systems ($\Omega_m < \kappa$), we see the nonlinearity is equally strong in the MIM and the single-cavity system. However, for sideband-resolved systems ($\Omega_m > \kappa$), the enhancement increases with sideband resolution factor. \autoref{fig:enhancement}d demonstrates that the MIM system can only feature larger quadratic transduction than in a single cavity when it is sideband-resolved. The absence of enhancement for a sideband-unresolved system ($\Omega_m \ll \kappa$) can be attributed to the fact that, in a single cavity, carrier, first and second sidebands are already resonantly enhanced due to the large spectral overlap.

	We derive an expression for the enhancement factor in the case of a sideband resolved system. We look at the case of constant $\bar{n}_{cav}$, driving of the even mode and finally large sideband resolution $\Omega_m\gg \kappa$ to simplify the expression. We find
	\begin{equation}\label{eq:sideband_limit}
	\left| \frac{\max_\Delta (\mathcal{A}^{(2)}_{e}(\Delta))}{\max_\Delta (\mathcal{A}^{(2)}_{ref}(\Delta))}\right| \approx 2\frac{\Omega_m}{\kappa},
	\end{equation}
	\autoref{eq:sideband_limit} demonstrates that, for a sideband-resolved system, the MIM system enhancement of nonlinearity is given by the degree of sideband resolution. This result is plotted as a red dashed line in figure \autoref{fig:enhancement}d. Similar results are obtained for an even input condition (not shown). 
	
	In \autoref{fig:enhancement}, we highlight the differences between mechanical transduction in a single cavity and in a MIM system. In \autoref{fig:enhancement}a, we see the characteristic MIM supermode frequency dependence on the static mechanical displacement $\bar{x}$. When static displacement is large (\autoref{fig:enhancement}b), the two cavities have frequencies that differ by more than $2J$ and we effectively recover the limit of two uncoupled cavities, whereas a zero static displacement gives the coupled cavity MIM system (\autoref{fig:enhancement} c). In \autoref{fig:enhancement}b and \autoref{fig:enhancement}c, we look at sidebands generated in a sideband-resolved MIM system corresponding to these crosscuts. For this, we assume a drive of the even mode and plot quantities $\mathcal{A}^{(1)}_o$ and $\mathcal{A}^{(2)}_e$. The horizontal dashed lines are the single cavity limits $\mathcal{A}_\mathrm{ref}^{(1)} \approx A_\mathrm{ref}^{(1)}$ and $\mathcal{A}_\mathrm{ref}^{(2)}\approx A_\mathrm{ref}^{(2)}$ for first and second sidebands as calculated previously. All plotted values are now normalised by $A_\mathrm{ref}^{(1)}$, which is done to give an idea of the relative size of first and second sidebands for currently available system parameters.
	
	In \autoref{fig:enhancement}b, we see that transduction for the uncoupled cavities adheres to the single cavity limits, as expected. Moving to the coupled cavity system in \autoref{fig:enhancement}c, we see that the second sideband amplitude now surpasses the single cavity limit. In \autoref{fig:enhancement}d, we plot the enhancement of the MIM system over a single cavity for second sideband amplitude: $\mathcal{A}^{(2)}/A_\mathrm{ref}^{(2)}$. We see that, at the double resonance condition, the enhancement peaks to the value of $2\Omega_m/\kappa$. 
	
	We predict that this effect could be experimentally observed in the currently available MIM systems, where sideband resolution reaches $\Omega_m/\kappa\approx 10$~\cite{Thompson2008,Sankey2010,Karuza2013}. Related coupled microtoroid resonators platforms~\cite{Grudinin2010} feature tunable inter-cavity coupling $J\leq\Omega_m/2$ and $\Omega_m/\kappa \approx 10$. An additional implementation of a coupled-cavity system was proposed for 2D optomechanical crystals~\cite{Safavi-Naeini2011a}, of which it was recently shown that individual cavities could reach $\Omega_m/\kappa \approx 28$~\cite{Ren2019}. 
	
	\subsection{Selectivity of quadratic over linear transduction}
	For experiments in which readout of the mechanical energy $\sim\hat{x}^2$ is desired, maximising the ratio of first to second-order sideband amplitude is crucial. This is because first sidebands carry information about $\hat{x}$ and their creation is thus inevitably associated with a linear quantum backaction that changes the mechanical state of the system~\cite{Miao2009,Yanay2016}.  
	
	As a figure of merit, we calculate the optimal ratio of the different sidebands, $\zeta = |A_{e,+}^{(2)}|/|A^{(1)}_{o,+}|$. From the equations (\autoref{fullsb2}), it can be derived this value is highest at the double resonance condition, which we find to be
	\begin{equation} \label{limit}
	\zeta \leq \frac{g_0 X_0}{\kappa}.
	\end{equation}
	Using \autoref{eq:singlecav_secondsb}, we easily see that this is the same limit as can be found in a single cavity. In other words, \autoref{limit} indicates that the MIM system does not allow for more selective generation of the second over first sideband as compared to a single cavity. In figure \autoref{fig:enhancement}e, we have plotted this sideband ratio as a function of $\Delta$ for $2J=\Omega_m$ and ground-state motion $X_0$=1. We see it also peaks at the double resonance condition, where it is limited by $g_0/\kappa$. We thus recover the condition found by Miao \textit{et al.}~\cite{Miao2009}, for a QND measurement of mechanical energy in the MIM system, in the ratio of sideband amplitudes, valid in either the classical or quantum domain and for general system parameters. As we will briefly discuss later, the calculation underlying \autoref{limit} is indeed closely related to an analysis of quantum measurement noise limits.
	
	Finally, we want to highlight another feature of the MIM system. Next to the (limited) enhancement of optomechanical nonlinearity, the MIM systems offer a simple method for separation of different sidebands, as they occur in orthogonal modes. Separation can be attained by a beam splitter (cf. \autoref{fig:fig1}a), even if the different sidebands are too close in frequency for the use of other filtering techniques. The degree of filtering this offers, though, is reduced when the cavity is not perfectly balanced, e.g. $g_{0,1} \neq g_{0,2}$ or $\kappa_1 \neq \kappa_2$, because the different sidebands are no longer output into orthogonal modes.
	
	\section{Back-action in the MIM system} \label{sec:apps}
	
	Having considered the effect of coherent mechanical motion on the cavity light field, we now move on to the effect of the light field on the resonator. In particular, we look at the well-known dynamical backaction (DBA) that occurs when the mechanically generated sidebands in the light field exert a force, whose sign and phase depends on laser detuning, back upon the resonator. Although these effects have been described in the MIM system previously ~\cite{Jayich2008,Heinrich2011,Lee2015,Paraiso2015}, we will now revisit these works using our general sideband picture to reinterpret and unify previous results.
	
	\subsection{Dynamical backaction and quadratic spring shift}
	Our approach starts again from the semiclassical equations of motion~\autoref{eq:opticalEOMS} and is similar to that of Jayich \textit{et al.}~\cite{Jayich2008}. A related method is used to determine DBA effects in single cavities~\cite{Aspelmeyer2014}. The aim is to find the susceptibility $\chi(\omega)$ of the mechanical resonator to an external force, given by the real amplitude $F_\mathrm{in}(t) = F_0\cos(\omega t)$. We solve for a mechanical motion that is strictly real, but can have an arbitrary phase that we account for by letting $X_0 \in \mathbb{C}$, i.e. $x(t) = (X_0 e^{i\omega t} + X_0^\ast e^{-i\omega t})/2$, . Note that this means information about both mechanical quadratures is now caught in the complex nature of $X_0$. We thus want to rewrite the mechanical EOM in the form $X_0(\omega) = \chi(\omega) F_0$. 
	
	For mechanical coherent motion given by $X_0$, we can write down the generated first sidebands using our previous \autoref{firstsb} and thus expand $|a_i|^2$ in \autoref{eq:opticalEOMS} in terms of $X_0$. In the present case we observe that the sidebands $A_{i,-}^{(1)}$ at $-\Omega_m$ actually depend on $X_0^\ast$ instead of $X_0$. By collecting all terms with the same time dependence, we can derive:
	\begin{subequations}
	\begin{align}
	\label{eq:susceptibility}
	\chi(\omega)^{-1} =& x_\mathrm{zpf} m\big[ -\omega^2 + \Omega_m^2 + i\Gamma_m \omega + \Omega_m (g_{0,1}\beta_{1,+},
	 - g_{0,2}\beta_{2,+}) \big],\\
	\beta_{i,+} =& \bar{a}_i \tilde{A}_{i,-}^\ast + \bar{a}_i^\ast \tilde{A}_{i,+},
	\end{align}
\end{subequations}
	and where $\tilde{A}_{i,-} = 2{A_{i,-}^{(1)}}/{X_0^\ast}$ and $\tilde{A}_{i,+} = 2{A_{i,+}^{(1)}}/{X_0}$.
	
	One of the striking features of a Hamiltonian with quadratic optomechanical coupling, as in \autoref{eq:effective_quad}, is that the optical cavity occupation $\bar{n}_c=\braket{\hat{a}_{c}^{\dagger}\hat{a}_{c}}$ directly changes the mechanical frequency by acting as an additional potential well for the resonator~\cite{Lee2015,Paraiso2015}. This can be seen from the Hamiltonian to be:
	\begin{equation} \label{statspringold}
	\Omega_\mathrm{eff} = \Omega_m + 2g_0^{(2)} \bar{n}_c.
	\end{equation}
	We shall refer to this effect as the static optical spring effect. Here, we show this effect can be described as a consequence of DBA, in which form it is much easier to include other DBA effects that can not be recovered from the quadratic coupling Hamiltonian, but are present in the MIM system. 
	
	By inserting $\Omega_\mathrm{eff} = \Omega_m + \delta \Omega$ and $\Gamma_\mathrm{eff} = \Gamma_m + \delta \Gamma$ into the susceptibility for $\bar{n}_c = 0$, and comparing to \autoref{eq:susceptibility}, we can find expressions for these shifts to be
	\begin{subequations} \label{eq:DBAshifts}
		\begin{align}
		\delta \Omega &= \frac{1}{2}\mathrm{Re}(g_{0,1}\beta_{1,+} - g_{0,2}\beta_{2,+}),\\
		\delta \Gamma_m &=  \mathrm{Im}(g_{0,1}\beta_{1,+} - g_{0,2}\beta_{2,+}).
		\end{align}
	\end{subequations}
	Now, we assume that the drive is close to resonance of the even supermode and, as before, that the two cavities are identical. In the adiabatic limit $2J\gg\Omega_m$, $\tilde{A}_{i,\pm}$ simplifies to $g_0\bar{a}_i/(2J)$ and $\beta_{1,+} \approx -\beta_{2,+}$. Combining these findings, we recover the quadratic coupling approximation: $\delta \Omega = g_0^2 \bar{n}_c/J$ by identifying $g_0^{(2)} = g_0^2/2J$. 
	
	\begin{figure}
		\centering
		\includegraphics[width=0.8\linewidth]{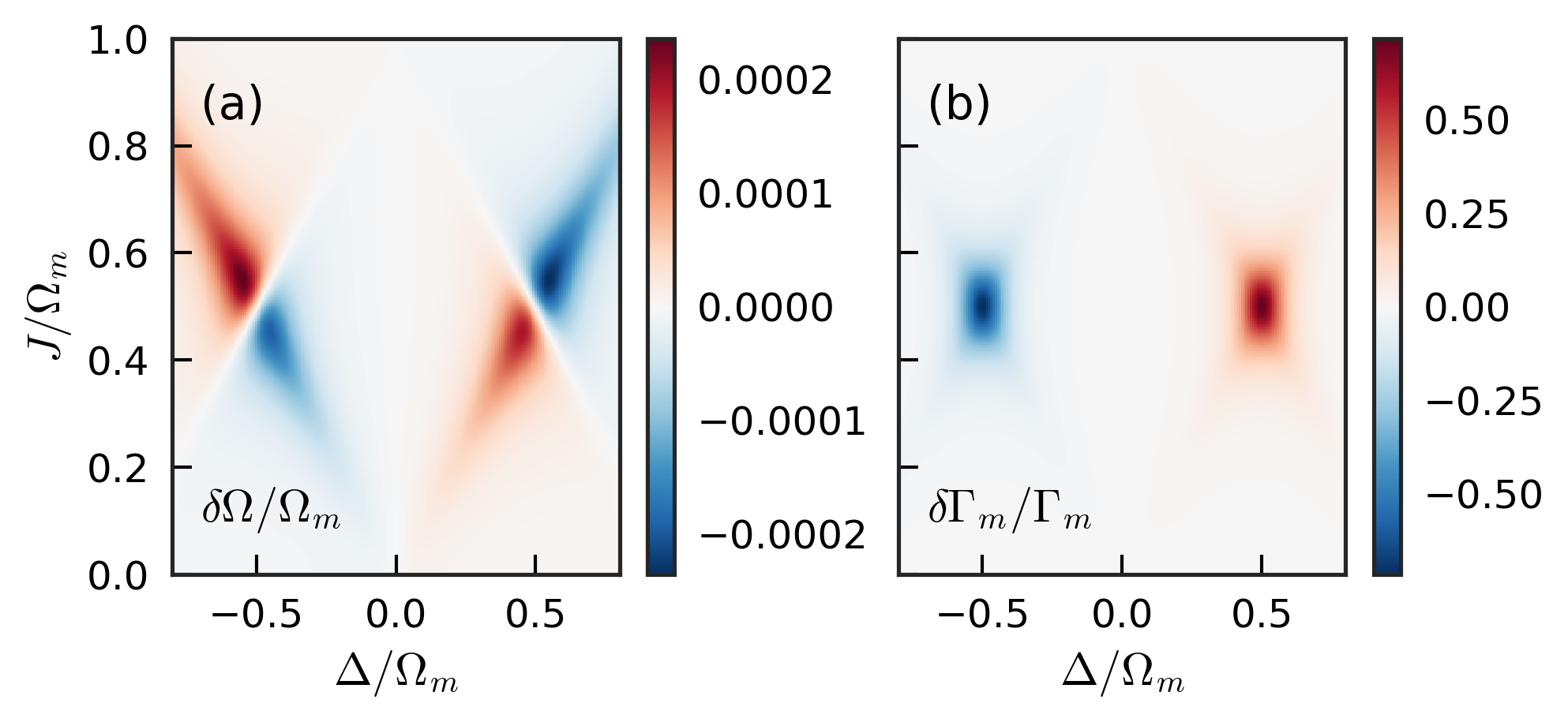}
		\caption{Dynamical back-action effects in the MIM system for an input power of 1 $\mu$W in cavity 1, $\frac{g_0}{2\pi} = 1$ MHz, $\frac{\Gamma_m}{2\pi} = 3$ MHz, $\frac{\kappa}{2\pi} = 1$ GHz and $\frac{\Omega_m}{2\pi} = 5$ GHz. (a) the optical spring effect normalised by the mechanical frequency (b) the optical amplification and cooling normalised by mechanical decay rate.}
		\label{fig:backaction}
	\end{figure}
	
	We see that the static optical spring effect can be regarded as a consequence of DBA, which considers only first sidebands, and thus is not a consequence of nonlinear optomechanical coupling. To be precise, for $J\gg\Omega_m$, the static optical spring effect is almost the same as the optical spring effect in a single cavity with a laser detuned from optical resonance by $J$. The only difference is that, due to the multimode MIM system, the carrier can be on resonance with one of the supermodes while the sidebands are far from resonance (i.e. the other supermode), allowing for the an optical spring effect with less input power, an idea related to that presented by Grudinin \textit{et al.}~\cite{Grudinin2010}. The reduction in input power is given by $\Delta_0/\kappa$, where $\Delta_0$ is the desired detuning from resonance for the particular application. To suppress unwanted DBA heating or cooling, it is generally taken larger than $\Omega_m$~\cite{Sonar2018}. This is one of the applications in which the MIM could outperform a single cavity: optical tuning the mechanical resonance through the optical spring effect using a detuned laser to suppress DBA heating or cooling in a sideband-resolved system.

	In order to further discuss optically-induced mechanical frequency and linewidth in the MIM system, we depict the relative modifications $\delta\Omega_m/\Omega_m$ and $\delta\Gamma_m/\Gamma_m$ as a function of $J$ and $\bar{\Delta}$ for constant input power \autoref{fig:backaction}. In \autoref{fig:backaction}a, we can see that, in the adiabatic regime $2J>\Omega_m$, we find the static optical spring effect around the supermode resonances, which closely resembles results from Lee \textit{et al.}~\cite{Lee2015}. Approaching the regime where $2J\approx\Omega_m$, the size of the optical spring increases because both sideband and carrier can be on resonant with one of the supermodes. A strong transition is found for $2J=\Omega_m$, where one of the sidebands crosses the resonance and changes the sign of the spring effect.  In \autoref{fig:backaction}b, we can see the optically-induced change in linewidth. The effect is again most substantial when both the carrier and one of the first sidebands are on resonance. Comparing to the standard optical spring effect, the linewidth change falls of more quickly when sidebands are not on resonance, which is already well known for single cavities~\cite{Aspelmeyer2014}.

	In previous work~\cite{Jayich2008}, Jayich \textit{et al.} extensively studied dynamical back-action as a function of inter-cavity detuning, here given by $\delta = \omega_{c,1}-\omega_{c,2}$. They noted a lack of backaction for $\delta=0$ in the adiabatic regime. It is  argued that backaction vanishes completely because of the fact that the first derivative of supermode frequency vanishes at $\delta=0$, suppressing linear coupling (see \autoref{fig:fig1}b). Here, however, we have seen that the DBA does not vanish completely. For $\delta=0$ and in the adiabatic regime, the first sideband amplitude is suppressed, as discussed in \autoref{sec:quasistatic}, but is not identically zero. This fact is important, because we have shown that second sideband amplitude (and thus nonlinear transduction) is suppressed when the first sideband amplitude is suppressed. Conversely, the generation of nonlinear transduction is associated with the presence of DBA. 
	
	This last statement can be seen as a classical analogue of previous results concerning the quantum non-demolition (QND) measurement of mechanical Fock states using quadratic optomechanical coupling~\cite{Miao2009,Yanay2016}. These authors showed that, as a result of the linear cross-mode coupling of the MIM system, the light field's vacuum fluctuations would destroy a mechanical Fock state before it could be measured through the effective $x^2$-coupling, unless the SPSC condition was fulfilled. An expression for quantum backaction was found by calculating the susceptibility of the optical modes to the input quantum fluctuations, leading to a result similar to $A^{(1)}_{i,\pm}$ in \autoref{firstsb}, where we calculate the susceptibility of the optical modes to mechanically-induced fluctuations. It is therefore also not surprising that we recover that the ratio of second to first sideband is limited by the same SPSC condition $g_0/\kappa$. Indeed, the ratio of second to first sideband amplitude is closely related to the ratio between the amount of information on $\hat{x}^2$ leaving the cavity and the quantum backaction, as the quantum backaction is directly related to the amount of information on $\hat{x}$ (i.e. the linear transduction) that leaves the cavity~\cite{Clerk2010a}.
	
	\subsection{Parametric squeezing}
	In parametric squeezing, the spring constant of a resonator is modulated at twice the mechanical frequency, which results in a quadrature-dependent amplification or damping of the resonator~\cite{Rugar1991}. Such a scheme has previously been used in electromechanical- (e.g.~\cite{Szorkovszky2013,Poot2014}) and linearly coupled optomechanical~\cite{Pontin2014,Sonar2018} systems. In a quadratically coupled optomechanical system, it is possible to directly alter the mechanical spring constant using the optical field, which can be exploited to implement this scheme~\cite{Nunnenkamp2010}. In fact, we find that the parametric squeezing effect lies at the heart of the two-phonon OMIT-like effect reported for the MIM system~\cite{Huang2011}. This can be seen from the fact that this OMIT effect works by amplifying thermal fluctuations in only one particular mechanical quadrature, de-amplifying motion in the opposite quadrature. We now set out to compare the parametric driving effect in the MIM system to a single cavity system.	 
	
\begin{figure} 
		\centering
		\includegraphics[width=\linewidth]{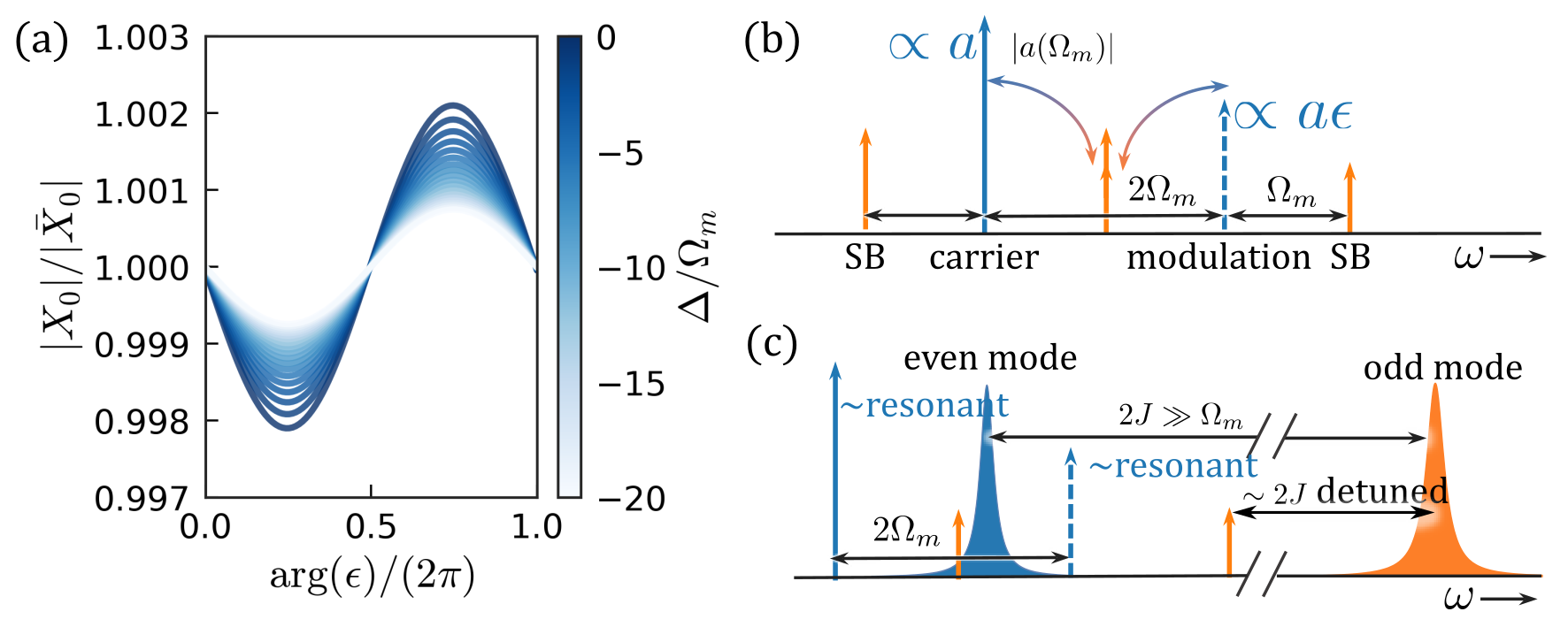}
		\caption{Parametric driving of the mechanical resonator using $2\Omega_m$-modulated light in the MIM system. (a) An example of thermal squeezing of the mechanical mode. By changing the phase between modulation tone $\epsilon$ and force $F_{ex}$, the response of the mechanical oscillator changes. For this plot, we assume a driven even mode with $\bar{n}_c = 1000$, $\frac{g_0}{2\pi} = 1$ MHz, $\frac{\Gamma_m}{2\pi} = 3$ MHz, $\frac{\kappa}{2\pi} = 0.25$ GHz, $\frac{\Omega_m}{2\pi} = 5$ GHz and large splitting $J=10\Omega_m$. (b) A schematic depiction of the parametric drive explained in terms of mechanically-generated sidebands of the carrier and modulation tone sidebands. The beating created by these sidebands acts back upon the mechanical resonator. (c) A proposed use of the MIM system in a parametric driving experiment. Exploiting the multimode character of the system, the required input power is reduced. In (b) and (c), the colours indicate whether light occupies the even (blue) or odd (orange) mode.}\label{fig:paramdrivefig}
	\end{figure}

	To include cavity modulation, we now start from an intracavity field given by:
	\begin{equation} \label{DBAinput}
	\bar{a}_i = a(1 + \epsilon e^{i2\Omega_mt}),
	\end{equation}
	where the constant $\epsilon\in\mathbb{C}$, assumed to be $|\epsilon| \ll 1$, controls the modulation phase and amplitude. Our approach share ingredients in common with that by Rugar and Grütter~\cite{Rugar1991}. We assume a force with fixed phase $F_{ex}(t) = F_0 \cos(\Omega_m t)$ and allow $x(t) = (X_0 e^{i\Omega t} + X_0^\ast e^{i\Omega t})/2$ as previously. The modulation sideband controlled by $\epsilon$ gives an additional component to $|a_i|^2(\pm \Omega)$, that shows up in \autoref{eq:opticalEOMS}. After making the dependence on $X_0$ explicit, the EOM from \autoref{eq:opticalEOMS} implies:
	\begin{multline} \label{eq:paramdriveEOM}
	\Omega_m\left[ i\Gamma_m  - (g_{0,1}\beta_{1,+} - g_{0,2}\beta_{2,+})\right]\frac{X_0}{2} = \left[ \Omega_m (g_{0,1}\beta_{1,-} - g_{0,2}\beta_{2,-}) \right]\frac{X_0^\ast}{2} + \frac{F_0}{2 x_\mathrm{zpf} m},
	\end{multline}
	where now $\beta_{i,-} = \bar{a}_i \epsilon \tilde{A}_{i,+}^\ast + \bar{a}_i^\ast E_{i}$ and the amplitudes for the sidebands generated from the modulation tone $a\epsilon$ by mechanical motion read
	\begin{subequations}
		\begin{align}
		E_1 &= -\epsilon\frac{-Jg_{0,2}\bar{a}_2 + (\bar{\Delta}_2 - \Omega_m)g_{0,1}\bar{a}_1}{(\bar{\Delta}_1 - \Omega_m)(\bar{\Delta}_2 - \Omega_m)-J^2}, \\
		E_2 &= \epsilon\frac{-Jg_{0,1}\bar{a}_1 + (\bar{\Delta}_1 - \Omega_m)g_{0,2}\bar{a}_2}{(\bar{\Delta}_1 - \Omega_m)(\bar{\Delta}_2 - \Omega_m)-J^2}.
		\end{align}
	\end{subequations}
	 Because the modulation tone is displaced by $2\Omega_m$ from the carrier, its sidebands have a different dependence on $\bar{\Delta}$ than the $\tilde{A}_{i,\pm}$. In \autoref{fig:paramdrivefig}b, we sketch the sidebands that are created and the associated contribution to the radiation pressure force. Here the carrier ($a$) and modulation tone sideband ($a\epsilon$) develop sidebands through mechanical motion. 
	
	Similarly, $X_0$ can be retrieved by combining \autoref{eq:paramdriveEOM} with its complex conjugate, to give
	\begin{equation} \label{eq:paramdrive_resp1}
	X_0 = \frac{c^\ast + d}{|c|^2 - |d|^2}\frac{F_0}{x_\mathrm{zpf} m}
	\end{equation}
	with
	\begin{equation} \label{eq:paramdrive_resp2}
	\begin{split}
	c &= i\Gamma_m\Omega_m - \Omega_m (g_{0,1}\beta_{1,+} - g_{0,2}\beta_{2,+}),\\
	d &= \Omega_m (g_{0,1}\beta_{1,-} - g_{0,2}\beta_{2,-}).
	\end{split}
	\end{equation}
	When changing the phase of $\epsilon$, $b$ changes with similar phase, altering $|X_0|$. This is quadrature-dependent amplification of motion: depending on the relative phase of the modulation tone $\epsilon$ and the force $F_{ex}$, the response $|X_0|$ of the system can be larger or smaller than in a system with no optomechanical coupling. In figure \autoref{fig:paramdrivefig}a, we have plotted an example of the mechanical response changing with the phase of $\epsilon$. 
	
	Now, from \autoref{eq:paramdrive_resp1}, we can make some observations on parametric driving in the MIM system. The amplitude of the enhanced mechanical quadrature depends on first sideband amplitudes, of which we have determined that these are not enhanced in the MIM system with respect to a single cavity for constant cavity number. In other words, although the MIM promises enhanced nonlinear coupling, the parametric drive per cavity photon is not larger than in a single cavity. 
	
	A system with multiple optical modes, such as the MIM, could, however, help to reduce the required input power, as was shown previously in the context of linear position measurement~\cite{Dobrindt2010} and phonon lasing~\cite{Grudinin2010}. Here, we propose a similar use that is particularly useful in optical parametric driving. A schematic example of the idea is shown in ~\autoref{fig:paramdrivefig}c. In an optomechanical parametric driving scheme, it is often desirable to have the carrier far detuned from the cavity to suppress DBA heating or cooling of the resonator~\cite{Sonar2018}. This means a considerable input power is needed to reach an appreciable intracavity photon number. In the application we envision, the carrier is on resonance with one of the two supermodes. In that case, the sidebands can be far off-resonant given that $2J\gg \Omega_m$, while requiring much less input power.
	
	\section{Heralded phonon pair generation}
	Previously the optomechanical interaction has been used in the heralded generation of single phonons~\cite{Galland2014,Hong2017}. When the optomechanical interaction is linearised through using a strong optical drive, Stokes scattering of a drive photon into the lower frequency sidebands is associated with the generation of a phonon. When using a mechanical system close to the ground state, the consecutive detection of a single Stokes photon within the mechanical decoherence time then heralds a 1-phonon mechanical Fock state. 
	
	Analogously, the detection of photons in a Stokes sideband shifted by $-2\Omega_m$ from the drive laser, created through a nonlinear optomechanical interaction, would herald the pairwise generation of two phonons. Specifically, if a single mechanical mode is involved, the detection heralds a 2-phonon Fock state in the resonator. This scheme works outside the SPSC regime. Here, we consider the feasibility of such a scheme in a MIM system, compare it to using a single cavity and discuss limitations due to the presence of first sideband photon generation.
	
	From the intracavity fields we calculated previously, we can calculate the output field by using the input-output relations~\cite{Aspelmeyer2014}. Assuming the input light field contains only carrier light, the output light field at the frequency of the first or second sideband is simply $\sqrt{\kappa_\mathrm{ex}} A^{(1,2)}_{i,\pm}$. Assuming optimal combination of the outputs of both cavities such that all photons in the proper cavity supermode are detected, the photon detection rate in any of the sidebands is
	\begin{equation}
	\Gamma_\pm^{(1,2)}=\kappa_\mathrm{ex}(|A^{(1,2)}_{1,\pm}|^2+|A^{(1,2)}_{2,\pm}|^2).
	\end{equation}
	We can evaluate the Stokes sidebands for a system initialised in the mechanical ground state by setting $X_0=2$ in equations \autoref{firstsb} and \autoref{fullsb2}, accounting for sideband asymmetry~\cite{Aspelmeyer2014}.
	
	We now consider a short measurement interval $\Delta t$ (which could be defined by the duration of an optical pulse)  and a low enough first sideband amplitude such that the probability $p_1 = \Delta t\Gamma_+^{(1)}$ of detecting a single photon in the first sideband is much smaller than unity, to ensure that a heralded state is not spoiled by a probabilistic excitation of single phonons. This condition sets an upper limit to the number of carrier photons that can be employed in a single measurement. We denote the maximum allowed probability of single-phonon generation (determined by the wanted level of purity) as $p_{1,max}$. With the associated maximum laser power, the probability $p_2=\Delta t\Gamma_+^{(2)}$ of detecting a photon in the second Stokes sideband to herald a pure two-phonon state is maximised at
	\begin{equation}
	p_2=\frac{p_2}{p_1}p_{1,max}\leq \left(\frac{g_0}{\kappa}\right)^2 p_{1,max},
	\end{equation}
	where we used our previous observation that $|A^{(2)}|/|A^{(1)}| \leq g_0/\kappa$.
	
	As we found before, this limitation holds for both single-cavity and MIM systems. Nonetheless, the optical power (intracavity photon number) that is required to reach the maximal rate of heralding two-phonon states is reduced for MIM systems at the optimal condition for second sideband generation, by a factor equal to $2\Omega_m/\kappa$, as we found in \autoref{eq:sideband_limit}. This leads to a practical advantage of the MIM system for this scheme, especially in cryogenic settings, where heating through laser absorption is often a significant limiting effect.
	
	\section{Conclusion}
	In this work, we have presented a general framework to describe nonlinear transduction and backaction effects in a MIM optomechanical system. Using this framework, we discuss in what applications a MIM system offers an advantage over an optomechanical cavity with single optical and mechanical mode. We show that the MIM system gives an enhancement of the intrinsic nonlinearity of the optomechanical interaction for supermode splitting $2J=\Omega_m$ that is limited by the degree of sideband resolution $\Omega_m/\kappa$. Additionally, the ratio of nonlinear to linear transduction in the MIM system is limited by the same condition as it is in the single cavity, namely $g_0/\kappa$, imposing constraints on the applications of the MIM system, as was previously shown for a QND measurement of phonon number~\cite{Miao2009}. In a discussion of backaction, we show that DBA in the MIM system is equal in strength per cavity photon to that in a single cavity, but is altered by the fact that the MIM system is multimode optically. Similarly, we discussed that a $2\Omega_m$-parametric driving scheme is also not enhanced in the MIM system, but that the multimode character of the system can be used to reduce the amount of input light required to reach a specific cavity photon number. Finally, we proposed a scheme to use the nonlinear interaction in the weak coupling regime to herald the generation of phonon pairs, for which we found that in the MIM system the required cavity photon number is reduced by $2\Omega_m/\kappa$ for a generation rate that is limited by the ratio of $g_0$ and $\kappa$.
	
	Although the above considerations all consider the MIM system, they can be applied to a larger class of multimode optomechanical systems. In several works that study quadratically-coupled optomechanical systems, second-order perturbation theory is used to derive the quadratic coupling coefficient from the unperturbed optical and mechanical mode fields~\cite{Rodriguez2011,Kaviani2015,Kalaee2016,Hauer2018}. The quadratic coupling coefficient $g_0^{(2)}$ is proportional to the second-order correction to the eigenmode frequency for a small perturbation of mechanical displacement:
	\begin{equation}
	g_0^{(2)} \propto \frac{\delta \omega^{(2)}}{\omega} \frac{1}{4}\frac{|\langle \mathbf{E}_\omega | \Delta \epsilon | \mathbf{E}_\omega\rangle|^2}{|\langle \mathbf{E}_\omega |\epsilon | \mathbf{E}_\omega\rangle|^2} -  \frac{1}{2}\sum_{\omega^\prime \neq \omega} \left( \frac{\omega^3}{\omega^{\prime 2} - \omega^2} \right) \frac{|\langle \mathbf{E}_{\omega^\prime} | \Delta \epsilon | \mathbf{E}_\omega\rangle|^2}{\langle \mathbf{E}_\omega |\epsilon | \mathbf{E}_\omega\rangle \langle \mathbf{E}_{\omega^\prime} |\epsilon | \mathbf{E}_{\omega^\prime}\rangle}.
	\end{equation}
	Here, $|\mathbf{E}_{\omega}\rangle$ indicates the electric field of a cavity eigenmode at frequency $\omega$, the bra-ket products indicate overlap integrals and $\Delta \epsilon,\delta\omega^{(2)}$ denote the change in system permittivity distribution $\epsilon$ and eigenfrequency, due to a small mechanical displacement $\Delta x$. In this equation, the first term is fully determined by, and much smaller than, $g_0$. The second term contains perturbation-induced overlaps between different eigenmodes, which are weighted by their frequency difference such that the contribution from closely spaced eigenmodes is enhanced. When applying this equation to the MIM system, it is the close spacing of $2J$ between the two supermodes that enhances quadratic coupling. However, it is this same mechanically-induced overlap between the two optical supermodes that gives the cross-mode optomechanical coupling, of which we have seen it limits the selectivity of quadratic over linear optomechanical coupling in the system. 
	
	At this point a question arises: given the generality of the second-order perturbation theory calculation, is it at all possible to design an optomechanical system such that it has a $x^2$-coupling without the linear cross-mode coupling? As already described by Miao \textit{et al.}~\cite{Miao2009}, any system that does have cross-coupling would always be restricted by the single-photon strong coupling requirement for QND measurements, and also be limited in that there will be residual linear DBA, as discussed in this paper. Currently, several proposals claim to circumvent this restriction~\cite{Kaviani2015,Hauer2018,Dellantonio2018a}. Although it is beyond the scope of this paper to discuss these works individually, the authors would like to stress that cross-coupling between any two modes may allow information about the position $x$ to escape the cavity and impose quantum backaction on the resonator.
	
	\section*{Acknowledgments}
	We would like to thank Pierre Busi, Andrea Fiore, Simon Gröblacher, Kevin Cognée and Femius Koenderink for valuable discussions. We thank Ilan Shlesinger for critical reading of the manuscript. This work is part of the research programme of the Netherlands Organisation for Scientific Research (NWO). E.V. acknowledges support from an NWO-Vidi grant and the European Research Council (ERC starting grant no. 759644-TOPP).

	\appendix
	\section{Homodyne signal in optimal quadrature in terms of sideband amplitudes} \label{appendixone}
	Consider homodyne detection on one of the two beamsplitter outputs from \autoref{fig:fig1} for even driving. Depending on the output, these contain either first- or second-order sidebands. We will assume first-order sidebands, although the exact same argument holds for second-order sidebands. The output of the beam splitter combined with a local oscillator field with amplitude $\bar{a}_\mathrm{L.O.}$ is given by the following expression
	\begin{equation}
	a_\mathrm{h.d.} = \bar{a}_\mathrm{L.O.}e^{i\theta} + \bar{a}_\mathrm{const} + \sqrt{\kappa_\mathrm{in}} A_{o,+}^{(1)} e^{i\Omega_m t} 
	+ \sqrt{\kappa_\mathrm{ex}} A_{o,-}^{(1)} e^{-i\Omega_m t}, 
	\end{equation}
	which we derived via the input-output relation $a_\mathrm{out} = a_\mathrm{in}-\sqrt{\kappa_\mathrm{in}}a$~\cite{Aspelmeyer2014}, under the assumption of large power $|\bar{a}_\mathrm{L.O.}\gg|\bar{a}_\mathrm{out}|$. Here  $\theta=\arg\bar{a}_\mathrm{L.O.}$ denotes the tunable local oscillator phase,  with $\bar{a}_\mathrm{const}$ containing all time-independent contributions to the output field and $A_{i,\pm}^{(1)}$ denoting the sideband amplitudes from \autoref{fullsb2}a,b. 
	
	The homodyne signal amplitude $S(\omega) \propto |a_\mathrm{h.d.}|^2(\omega)$ at frequency $\Omega_m$ is found to be
	\begin{align}
	S(\Omega_m)\propto&
	\sqrt{\kappa_\mathrm{ex}}\bar{a}_\mathrm{L.O.}\big[e^{i\theta}(A_{o,+}^{(1)\ast}e^{-i\Omega_m t} + A_{o,-}^{(1)\ast} e^{i\Omega_m t}) 
	+ e^{-i\theta}(A_{o,+}^{(1)} e^{i\Omega_m t} + A_{o,-}^{(1)} e^{-i\Omega_m t})\big] \nonumber\\
	&\simeq 2\sqrt{\kappa_\mathrm{ex}}\bar{a}_\mathrm{L.O.} \mathrm{Re}[e^{i\theta}B(t)],
	\end{align}
	where $B(t) = A_{o,+}^{(1)\ast}e^{-i\Omega_m t} + A_{o,-}^{(1)\ast} e^{i\Omega_m t}$ and we have only slowly-oscillating terms. To optimise homodyne signal, we set $\theta$ such that, for $|B_\mathrm{max}| = \max_t (|B(t)|)$ and $B_\mathrm{max}$ the corresponding complex value, $e^{i\theta}B_\mathrm{max}$ is real. We then find that $S(\Omega_m) \propto \sqrt{\kappa_\mathrm{ex}} \bar{a}_\mathrm{L.O.} |B_\mathrm{max}|$.  Given that $B(t)$ is the sum of two counterrotating complex amplitudes, its norm is largest when these have the same phase, thus if $|B_\mathrm{max}| = |A_{o,-}^{(1)}| + |A_{o,+}^{(1)}|$ and 
	\begin{equation}
	S(\Omega_m)  \propto \sqrt{\kappa_\mathrm{ex}} \bar{a}_\mathrm{L.O.} \left(|A_{o,-}^{(1)}| + |A_{o,+}^{(1)}| \right).
	\end{equation}
	This derivation demonstrates the metric we use is a measure of the signal amplitude in the optimal homodyne measurement.

\newcommand\newblock{\hskip .11em plus .33em minus .07em} 
\def\bibfont{\footnotesize}

\bibliographystyle{apsrev4-1}
\bibliography{njpbibopts,library}

\end{document}